# Strong quantum entanglement via a controllable four-wave-mixing mechanism in an optomechanical system

Xiang You[1, 2] and Yongmin Li[1, 2,*]

[1]State Key Laboratory of Quantum Optics and Quantum Optics Devices, Institute of Opto-Electronics, Shanxi University, Taiyuan 030006, People's Republic of China
[2]Collaborative Innovation Center of Extreme Optics, Shanxi University, Taiyuan 030006, People's Republic of China

We propose an approach to generate strong quantum entanglement by the controllable four-wave-mixing mechanism in a single-cavity, weak-coupling optomechanical system. The optomechanical system is driven by a strong two-tone pump field and a weak signal field, simultaneously. The two-tone pump field consists of a lower and an upper sideband, which couple with the optical cavity and mechanical resonator, and generate the beam-splitter and two-mode squeezing interactions under the rotating-wave approximation. This interaction mechanism modifies the effective susceptibility of the optomechanical cavity and optomechanically induces a four-wave-mixing process. Strong quantum entanglement can be generated between the signal and four-wave-mixing fields with an entangled degree over 16 dB in realistic optomechanical systems. The generation scheme of the quantum entanglement is quite robust against thermal–mechanical noise, and entanglement above 3 dB can persist at room temperature in the weak-coupling regime.

## I. INTRODUCTION

Typical optomechanical first-order interactions include the beam-splitter (BS) and two-mode squeezing (TMS) interactions [1], which can produce various physical phenomena. The BS interaction enables the exchange between photons and phonons that contribute to sideband cooling [2-7], coherent quantum state transfer, wavelength conversion [8-12], and optomechanically induced transparency [13, 14]. In comparison, the TMS interaction enables the generation of photon–phonon pairs, and has been used to realize the entanglement between photons and phonons [15] and optomechanically induced amplification [16]. By combining the BS and TMS interactions, the Bogoliubov mode in the optomechanical system can be generated [17-19]. The back-action evasion measurement for the mechanical quadrature component was achieved when the coupling strengths of the BS and TMS interactions were equal [20-23]. Moreover, the quantum squeezed state of the mechanical mode has been generated by increasing the coupling strength of the BS interaction to be larger than that of the TMS interaction [18, 24-28].

Squeezed light fields are typical non-classical states where the fluctuation noise in one quadrature of the optical field is below the standard quantum noise limit; these fields have been demonstrated in optomechanical systems [29-31]. The two-mode squeezed state is a special type of quantum entangled state, where the amplitude quadrature and phase quadrature of the states are both quantum-correlated. Quantum entanglement is a crucial resource for quantum communication and quantum computing [32, 33], it can improve measurement sensitivity, and is beneficial to quantum metrology [34, 35]. The quantum entanglement of two output optical fields in optomechanical systems has been investigated by coupling a cavity mode with a mechanical mode [36,37] and by coupling two cavity modes simultaneously with a mechanical mode [38,39].

In this study, we present a scheme to produce strong quantum entanglement via the controllable four-wave-mixing (FWM) process in a single-cavity, weak-coupling optomechanical system. Our scheme requires neither the strong coupling condition nor multiple cavity modes. Instead, only a two-tone pump field is required, which couples with the mechanical resonator to generate the desired BS and TMS interactions. In this way, the effective susceptibility of the optomechanical cavity is modified, and an optomechanically FWM process is induced when a weak signal field is incident on the system.

The FWM enables a significant amplification of the signal field and generates an associated FWM field. We find that there exists remarkable quantum correlation for the amplitude (phase) quadrature sum (difference) between the signal and the FWM fields with realistic experimental parameters, that is, strong quantum entanglement can be generated. Our scheme works in a resolved-sideband regime, which facilitates the high-efficiency laser-cooling of the mechanical mode, a key factor for achieving strong quantum entanglement for optomechanical systems. We show that even without the strong optomechanical coupling, considerable quantum entanglement at room temperature can still be achieved by carefully controlling the ratio of BS and TMS interactions.

The rest of this study is organized as follows: In Section II, we present the theoretical model of the optomechanical system and explicitly derive the output field, which consists of the classical field and quantum fluctuation field, by solving the quantum Langevin equation. In Section III, we investigate the classical characteristics of the signal and FWM fields, including the bandwidth, center frequency, and intensity gain. In Section IV, we study the quantum entanglement between the signal and FWM fields, and analyze the dependence of the entanglement on the relevant experimental parameters, including the optomechanical coupling strength, escape efficiency of optical cavity, and environmental temperature. Finally, conclusions are given in Section V.

## II. OPTOMECHANICAL SYSTEM MODEL

### A. Hamiltonian and Langevin equations

We consider an optomechanical system consisting of a mechanical resonator with resonance frequency $\omega_m$ and an optical cavity with intrinsic resonance frequency $\omega_{c0}$. The optical cavity is driven by an intense two-tone pump field consisting of a red-shifted sideband $\omega_- = \omega_c - \omega_0$, a blue-shifted sideband $\omega_+ = \omega_c + \omega_0$, and a weak probe field with frequency $\omega_s$. Here, $\omega_c$ is the effective resonance frequency of the optical cavity considering the average radiation pressure of the two-tone pump field. As we will show below, in the resolved-sideband regime $\kappa \ll \omega_m$ and under the rotating-wave approximation, the optomechanical coupling of the red-shifted and blue-shifted sidebands with the mechanical resonator produce the BS interaction and TMS interaction, respectively.

The Hamiltonian of the optomechanical system is written as

$$H = \hbar\omega_{c0} a^\dagger a + \hbar\omega_m b^\dagger b + \hbar g_0 a^\dagger a \left(b^\dagger + b\right) + H_{\text{drive}}, \qquad (1)$$

where $a(b)$ is the photon (phonon) annihilation operator, and $g_0$ denotes the single-photon optomechanical coupling strength. The first two terms on the right-hand side of Eq. (1) are the free Hamiltonian of the cavity field and the mechanical mode, respectively. The third term is the interaction Hamiltonian, and $H_{\text{drive}}$ denotes the driving Hamiltonian, which is written as

$$H_{\text{drive}} = i\hbar\sqrt{\kappa_{ex}} \left(\alpha_+ e^{-i\omega_+ t} + \alpha_- e^{-i\omega_- t} + \alpha_s e^{-i\omega_s t}\right) a^\dagger + \text{H.c.}, \qquad (2)$$

where $\kappa_{ex}$ is the decay rate of the input cavity mirror (external loss rate), and $\kappa_0$ is the internal loss rate of the cavity apart from $\kappa_{ex}$, which results in a total decay rate $\kappa = \kappa_{ex} + \kappa_0$. The driving strength $|\alpha_\mu| = \sqrt{P_\mu/\hbar\omega_\mu}$ $(\mu = \pm, s)$ is related to the input laser power $P_\mu$, and we assume that $|\alpha_+|, |\alpha_-| \gg |\alpha_s|$.

The quantum Langevin equations of the cavity field $a$ and mechanical mode $b$ have the forms

$$\dot{a} = -(i\omega_{c0} + \kappa/2)a - ig_0 a(b + b^\dagger) + \sqrt{\kappa_{ex}}\varepsilon(t) + \sqrt{\kappa_{ex}} a_{in} + \sqrt{\kappa_0} a_v, \qquad (3)$$

$$\dot{b} = -(i\omega_m + \gamma_m/2)b - ig_0 a^\dagger a + \sqrt{\gamma_m}\eta, \qquad (4)$$

where $\varepsilon(t) = \alpha_+ e^{-i\omega_+ t} + \alpha_- e^{-i\omega_- t} + \alpha_s e^{-i\omega_s t}$ denotes the total driving fields, $a_{in}$ and $a_v$ denote the quantum fluctuation noise of the input field and vacuum field, respectively.

$\gamma_m$ is the intrinsic mechanical damping rate, and $\eta$ is the thermal drive to the mechanical resonator.

We apply the transformations $a = \bar{\alpha}_- e^{-i\omega_- t} + \bar{\alpha}_+ e^{-i\omega_+ t} + a_1$ and $b = \bar{\beta} + b_1$ to the optical field and mechanical mode, where $a_1$ and $b_1$ represent the fluctuation fields of the cavity mode $a$ and mechanical mode $b$, respectively, and $\bar{\alpha}_\pm = \sqrt{\kappa_{ex}}\alpha_\pm / (\mp i\omega_0 + \kappa/2)$ represents the mean intracavity coherent amplitude of the two-tone pump field. Without loss of generality, we assume $\bar{\alpha}_\pm$ to be real in the following. Note that the intrinsic resonance frequency of the optical cavity $\omega_{c0}$ is shifted to $\omega_c = \omega_{c0} + g_0 \bar{x}$ by the average displacement of the mechanical mode, which is given by $\bar{x} = \bar{\beta} + \bar{\beta}^* = -2g_0(\bar{\alpha}_-^2 + \bar{\alpha}_+^2)/\omega_m$.

In the resolved-sideband regime $\kappa \ll \omega_m$, by transforming the system to a rotating frame defined by $H_0 = \hbar\omega_c a_1^\dagger a_1 + \hbar\omega_0 b_1^\dagger b_1$ and applying a rotating-wave approximation, the linearized interaction Hamiltonian can be written as follows by neglecting the nonlinear terms:

$$H = \hbar\Delta_m b_1^\dagger b_1 + \hbar G_-\left(a_1^\dagger b_1 + a_1 b_1^\dagger\right) + \hbar G_+\left(a_1^\dagger b_1^\dagger + a_1 b_1\right), \tag{5}$$

where $\Delta_m = \omega_m - \omega_0$ denotes the mechanical frequency detuning between the intrinsic mechanical frequency $\omega_m$ and the modulated frequency $\omega_0$, and $G_- = g_0 \bar{\alpha}_-$ ($G_+ = g_0 \bar{\alpha}_+$) denotes the optomechanical coupling strength of the BS (TMS) interaction arising from the red-shifted (blue-shifted) sideband. Starting with Eq. (5), the quantum Langevin equations of the fluctuation fields $a_1$ and $b_1$ are given by

$$\dot{a}_1 = (-\kappa/2)a_1 - i\left(G_- b_1 + G_+ b_1^\dagger\right) + \sqrt{\kappa_{ex}}\alpha_s e^{-i\Delta_s t} + \sqrt{\kappa_{ex}}a_{in} + \sqrt{\kappa_0}a_v, \tag{6}$$

$$\dot{b}_1 = -(i\Delta_m + \gamma_m/2)b_1 - i\left(G_- a_1 + G_+ a_1^\dagger\right) + \sqrt{\gamma_m}\eta, \tag{7}$$

where $\Delta_s = \omega_s - \omega_c$ is the frequency difference between the signal field frequency $\omega_s$ and the effective resonance frequency of the optical cavity $\omega_c$.

## B. Solutions in the frequency domain

Using the Fourier transform $o[\omega] = \int_{-\infty}^{\infty} o(t)e^{i\omega t}dt$, Eqs. (6) and (7) are converted into the frequency domain:

$$a_1[\omega] = \chi_c[\omega]\left\{-i\left(G_-b_1[\omega]+G_+b_1^\dagger[\omega]\right)+\sqrt{\kappa_{ex}}\alpha_s[\omega-\Delta_s]+\sqrt{\kappa_{ex}}a_{in}[\omega]+\sqrt{\kappa_0}a_v[\omega]\right\}, \quad (8)$$

$$b_1[\omega] = \chi_m[\omega]\left\{-i\left(G_-a_1[\omega]+G_+a_1^\dagger[\omega]\right)+\sqrt{\gamma_m}\eta[\omega]\right\}, \quad (9)$$

where $\chi_c[\omega] = -(i\omega+\kappa/2)^{-1}$ and $\chi_m[\omega] = -[(i\omega-\Delta_m)+\gamma_m/2]^{-1}$ are the susceptibilities of the optical cavity and mechanical resonator, respectively. Combining Eqs. (8) and (9), and using the operator conjugate relation $(o[\omega])^\dagger = o^\dagger[-\omega]$, we obtain the fluctuation field of the optical cavity mode in the frequency domain

$$\begin{aligned}a_1[\omega] = &A[\omega]\left(\sqrt{\kappa_{ex}}\alpha_s[\omega-\Delta_s]+\sqrt{\kappa_{ex}}a_{in}[\omega]+\sqrt{\kappa_0}a_v[\omega]\right)\\ &+B[\omega]\left(\sqrt{\kappa_{ex}}\alpha_s^*[\omega+\Delta_s]+\sqrt{\kappa_{ex}}a_{in}^\dagger[\omega]+\sqrt{\kappa_0}a_v^\dagger[\omega]\right).\\ &+C[\omega]\sqrt{\gamma_m}\eta[\omega]+D[\omega]\sqrt{\gamma_m}\eta^\dagger[\omega]\end{aligned} \quad (10)$$

The coefficients that depend on the related parameters of the optomechanical system are given by

$$\begin{aligned}A[\omega] &= \frac{\chi_c[\omega]\left\{1+\chi_c[\omega]\left(G_-^2\chi_m^*[-\omega]-G_+^2\chi_m[\omega]\right)\right\}}{(1+\chi_m[\omega]\Sigma[\omega])(1+\chi_m^*[-\omega]\Sigma[\omega])},\\ B[\omega] &= \frac{\chi_c[\omega]\chi_c[\omega]G_-G_+\left(\chi_m^*[-\omega]-\chi_m[\omega]\right)}{(1+\chi_m[\omega]\Sigma[\omega])(1+\chi_m^*[-\omega]\Sigma[\omega])},\\ C[\omega] &= -i\frac{\chi_c[\omega]\chi_m[\omega]G_-}{(1+\chi_m[\omega]\Sigma[\omega])},\\ D[\omega] &= -i\frac{\chi_c[\omega]\chi_m^*[-\omega]G_+}{(1+\chi_m^*[-\omega]\Sigma[\omega])}.\end{aligned} \quad (11)$$

Here, $\Sigma[\omega]$ represents the optomechanical self-energy that is derived from the unequal optomechanical coupling strength between the BS and TMS interactions, and is defined as

$$\Sigma[\omega] = \chi_c[\omega]\left(G_-^2-G_+^2\right). \quad (12)$$

The real and imaginary parts of the optomechanical self-energy correspond to the frequency-dependent optomechanical damping rate $\gamma_{opt} = 2\operatorname{Re}(\Sigma[\omega])$ and mechanical frequency shift $\delta\omega_m[\omega] = \operatorname{Im}(\Sigma[\omega])$, respectively. In the weak-coupling regime $G_\pm \ll \kappa$, the corresponding optomechanical damping rate and mechanical frequency shift [1] are given by

$$\gamma_{opt} = \frac{\kappa\left(G_-^2 - G_+^2\right)}{\Delta_m^2 + \kappa^2/4}, \quad (13)$$

$$\delta\omega_m = \frac{\Delta_m\left(G_-^2 - G_+^2\right)}{\Delta_m^2 + \kappa^2/4}. \quad (14)$$

Inserting Eq. (10) into the input–output relation $a_{out}[\omega] = -a_{in}[\omega] + \sqrt{\kappa_{ex}}a_1[\omega]$, we obtain the reflective field from the optical cavity:

$$\begin{aligned}
a_r[\omega] &= \left(A[\omega]\kappa_{ex} - 1\right)\left(\alpha_s[\omega - \Delta_s] + a_{in}[\omega]\right) + A[\omega]\sqrt{\kappa_{ex}\kappa_0}a_v[\omega] \\
&+ B[\omega]\kappa_{ex}\left(\alpha_s^*[\omega + \Delta_s] + a_{in}^\dagger[\omega]\right) + B[\omega]\sqrt{\kappa_{ex}\kappa_0}a_v^\dagger[\omega] \\
&+ C[\omega]\sqrt{\kappa_{ex}\gamma_m}\eta[\omega] + D[\omega]\sqrt{\kappa_{ex}\gamma_m}\eta^\dagger[\omega]
\end{aligned} \quad (15)$$

The reflective field $a_r[\omega]$ consists of two parts: the classical field $\alpha_r$ and the quantum fluctuation field $a_{r1}$, which are expressed by

$$\alpha_r[\omega] = \left(A[\omega]\kappa_{ex} - 1\right)\alpha_s[\omega - \Delta_s] + B[\omega]\kappa_{ex}\alpha_s^*[\omega + \Delta_s], \quad (16)$$

$$\begin{aligned}
a_{r1}[\omega] &= \left(A[\omega]\kappa_{ex} - 1\right)a_{in}[\omega] + A[\omega]\sqrt{\kappa_{ex}\kappa_0}a_v[\omega] + B[\omega]\kappa_{ex}a_{in}^\dagger[\omega] \\
&+ B[\omega]\sqrt{\kappa_{ex}\kappa_0}a_v^\dagger[\omega] + C[\omega]\sqrt{\kappa_{ex}\gamma_m}\eta[\omega] + D[\omega]\sqrt{\kappa_{ex}\gamma_m}\eta^\dagger[\omega]
\end{aligned}. \quad (17)$$

From Eqs. (16) and (17), it is clear that the optomechanical interaction mechanism modifies the susceptibility of the optical cavity from its original form $\chi_c[\omega]$ and induces a conjugate field $\alpha_s^*[\omega + \Delta_s]$ (FWM field). The modified susceptibility of the optical cavity depends on $A[\omega]$, $B[\omega]$, $C[\omega]$, and $D[\omega]$, which in turn rely on the parameters of the optomechanical system.

### III. OPTOMECHANICAL FWM PROCESS

In this section, we investigate the frequency response characteristics of the classical field. We assume that the input signal field $\alpha_s$ is a monochromatic field with the form $\alpha_s[\omega] = \alpha_s\delta[\omega - \Delta_s]$. From Eq. (16), there are two frequency components $\omega = \Delta_s$ and $\omega = -\Delta_s$ in the reflective classical field $\alpha_r[\omega]$:

$$\alpha_{r-s}[\Delta_s] = (A[\Delta_s]\kappa_{ex} - 1)\alpha_s, \tag{18}$$

$$\alpha_{r-c}[-\Delta_s] = B[-\Delta_s]\kappa_{ex}\alpha_s^*, \tag{19}$$

where $\alpha_{r-s}[\Delta_s]$ and $\alpha_{r-c}[-\Delta_s]$ correspond to the signal and FWM fields ($\omega_F = 2\omega_c - \omega_s$), respectively, as shown in Fig. 1. From the above equations, the intensity gain of the reflective signal and FWM fields are

$$R_s[\Delta_s] = |\alpha_{r-s}[\Delta_s]/\alpha_s|^2 = |A[\Delta_s]\kappa_{ex} - 1|^2, \tag{20}$$

$$R_c[-\Delta_s] = |\alpha_{r-c}[-\Delta_s]/\alpha_s^*|^2 = |B[-\Delta_s]\kappa_{ex}|^2. \tag{21}$$

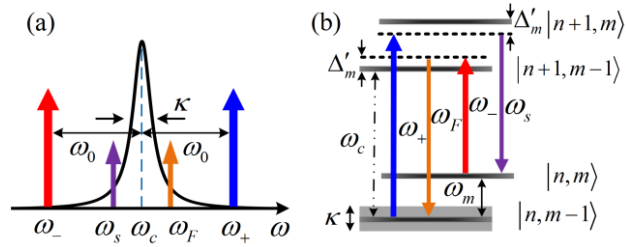

Fig. 1. (a) Spectrum of cavity response, optical driving fields, and FWM fields. $\omega_c$, effective cavity resonance frequency; $\omega_-$ and $\omega_+$, two-tone driving field; $\omega_s$, signal field; $\omega_F$, FWM field. (b) Energy-level diagram of the optomechanical system. $\Delta'_m$ is the effective mechanical detuning, $m$ and $n$ denote the phonon number of the mechanical mode and the photon number of the cavity mode, respectively.

Figure 2 plots the intensity gain $R_s[\Delta_s]$ ($R_c[-\Delta_s]$) of the signal (FWM) field as a function of the frequency detuning $\Delta_s$. The simulation parameters we use are $\omega_m = 2\pi \times 5.85 \times 10^5$ Hz, $\gamma_m = 2\pi \times 5$ Hz, $\kappa = 0.1\,\omega_m$, $\kappa_{ex} = 0.98\,\kappa$, and $\omega_0 = 0.95\,\omega_m$. The curves in Fig. 2 correspond to the TMS interaction coupling strength of $G_+ = \{30070\text{ Hz}, 30000\text{ Hz}, 29900\text{ Hz}\}$ from top to bottom, respectively, in which the BS interaction strength is set as $G_- = 3 \times 10^4$ Hz. For all curves, the weak-coupling condition $G_\pm \ll \kappa$ is satisfied. The signal (FWM) field presents a super-narrow frequency response bandwidth determined by the mechanical effective damping rate $\gamma_{eff} = \gamma_m + \gamma_{opt}$. The peak intensity gain of the signal (FWM) field occurs at the frequency of $\Delta_s = \mp\Delta'_m$ $(-\Delta_s = \pm\Delta'_m)$, which indicates that injection of the weak signal field with frequency $\Delta_s = \mp\Delta'_m$ into the optomechanical system induces a FWM field with frequency $-\Delta_s = \pm\Delta'_m$ when the strong two-tone pump field is applied, as shown in Fig. 1(b). Note that $\Delta'_m = \Delta_m + \delta\omega_m$ denotes the effective mechanical frequency detuning, considering the mechanical frequency shift induced by the optomechanical self-energy (see Eq. (14)).

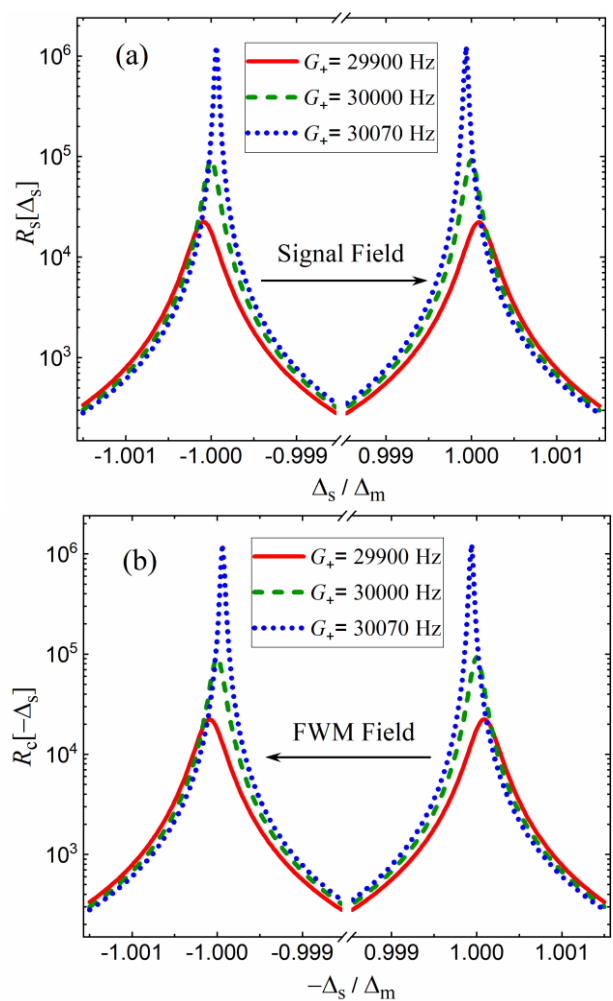

Fig. 2. Intensity gain spectrum of the reflective field versus the signal frequency detuning $\Delta_s$ for different coupling strengths: (a) signal field and (b) FWM field.

The frequency response characteristics of the FWM process is tunable by varying the optomechanical self-energy. In Fig. 3, we plot the effective mechanical frequency detuning $\Delta'_m$ and effective damping rate $\gamma_{eff}$ as a function of the coupling strength

$G_+$. Other simulation parameters are the same as those in Fig. 2. The circles A, B, and C represent three different $G_+$ in Fig. 2 with A: $G_+ = 29900$ Hz, B: $G_+ = 30000$ Hz, and C: $G_+ = 30070$ Hz. Note that the TMS interaction strength should satisfy $G_+ < 30100$ Hz (point D) to avoid the instability effect of the mechanical system.

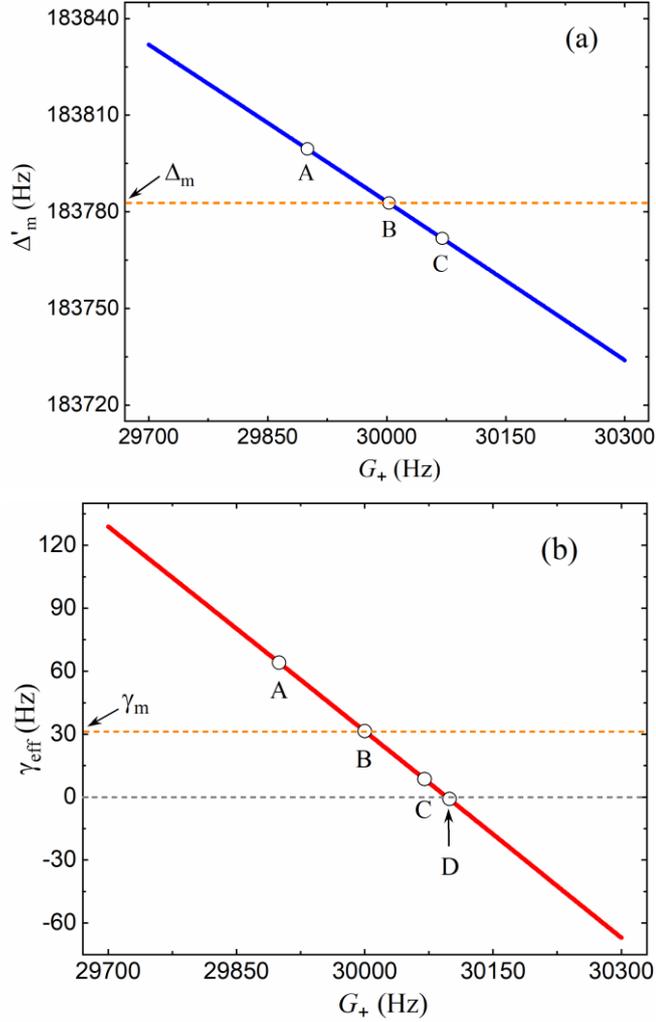

Fig. 3. (a) Effective mechanical frequency detuning $\Delta'_m$ and (b) effective damping rate $\gamma_{eff}$ as a function of the TMS interaction strength $G_+$. For all cases, $G_- = 3 \times 10^4$ Hz.

From Eqs. (20) and (21), the peak intensity gain of the signal (FWM) field at the center frequency $\Delta_s = -\Delta'_m$ ($-\Delta_s = \Delta'_m$) is

$$R_s = \left| A[-\Delta'_m] \kappa_{ex} - 1 \right|^2, \tag{22}$$

$$R_c = \left| B[\Delta'_m] \kappa_{ex} \right|^2. \tag{23}$$

Figure 4 shows the peak intensity gain of the signal (FWM) field as a function of the mechanical intrinsic damping rate $\gamma_m$, in which the optomechanical self-energy is set to zero, that is, $G_+ = G_- = 3\times10^4$ Hz, and the other simulation parameters are the same as those in Fig. 2. We find that the lower mechanical intrinsic damping rate $\gamma_m$ makes a higher intensity gain, which can exceed $10^5$ at the region of $\gamma_m < 30$ Hz. This indicates that a high-quality factor of the mechanical resonator predicts a high intensity gain in the absence of optomechanical self-energy.

For a given mechanical resonator with a fixed quality factor, one can tune the peak intensity gain by controlling the optomechanical self-energy. To this end, we fix $G_-$ and adjust $G_+$ to 29900 Hz (point A), 30000 Hz (point B), and 30070 Hz (point C), as shown in Fig. 3(b), the effective mechanical damping rate is changed to $\gamma_{\text{eff}} = \{64.0 \text{ Hz}, 31.4 \text{ Hz}, 8.5 \text{ Hz}\}$. In order to verify that the effective mechanical damping $\gamma_{\text{eff}}$ has the same tuning effect on the intensity gain as the intrinsic damping rate $\gamma_m$, we extract three intrinsic damping rates: A′: 8.5 Hz, B′: 31.4 Hz, and C′: 64.0 Hz. These are presented in Fig. 4, and have the same values as A, B, and C. We find that the intensity gains of A′, B′, and C′ in Fig. 4 are equal to the peak gains of A, B, and C, respectively (see Fig. 2). According to the above analysis, the peak gain of the intensity mainly depends on the effective mechanical damping rate when the TMS coupling strength $G_+$ varies around $G_-$.

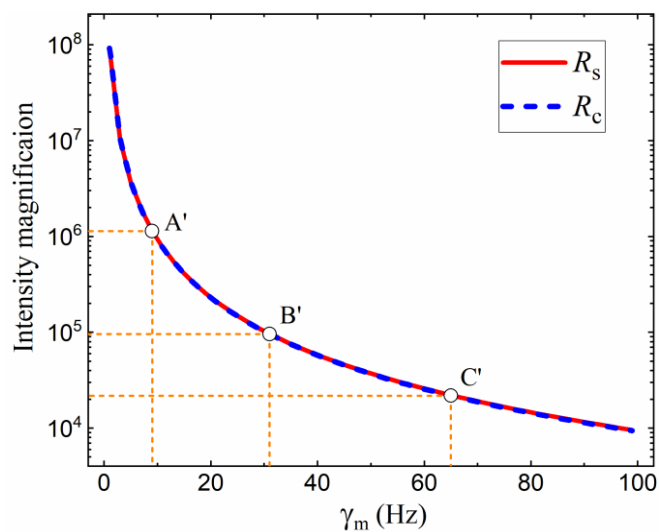

Fig. 4. Intensity gain of the signal and FWM fields as a function of the mechanical intrinsic damping rate $\gamma_m$.

## IV. THE QUANTUM ENTANGLEMENT

In this section, we study the phenomenon of quantum entanglement between the signal and FWM fields. The amplitude quadrature and phase quadrature of the signal and FWM fields in the frequency domain are defined as

$$X_s[\omega] = \left(a_{r1}[\omega - \Delta_s] + a_{r1}^\dagger[\omega + \Delta_s]\right)/\sqrt{2}, \tag{24}$$

$$Y_s[\omega] = -i\left(a_{r1}[\omega - \Delta_s] - a_{r1}^\dagger[\omega + \Delta_s]\right)/\sqrt{2}, \tag{25}$$

$$X_c[\omega] = \left(a_{r1}[\omega + \Delta_s] + a_{r1}^\dagger[\omega - \Delta_s]\right)/\sqrt{2}, \tag{26}$$

$$Y_c[\omega] = -i\left(a_{r1}[\omega + \Delta_s] - a_{r1}^\dagger[\omega - \Delta_s]\right)/\sqrt{2}. \tag{27}$$

From the above optical field quadratures, we further define two combined field quadratures, including both signal and FWM fields. These are written as

$$X^+[\omega] = \left(X_s[\omega] + X_c[\omega]\right)/\sqrt{2}, \tag{28}$$

$$Y^-[\omega] = \left(Y_s[\omega] - Y_c[\omega]\right)/\sqrt{2}. \tag{29}$$

Using the definition of the noise power spectrum $S_{xx}[\omega] = \int_{-\infty}^{\infty} \langle x(\omega) x(\omega') \rangle d\omega'$, the noise power spectrum of the combined field quadratures, more precisely, the amplitude quadrature sum noise power spectrum $S_{XX}^+[\omega]$ and the phase quadrature difference noise power spectrum $S_{YY}^-[\omega]$, can be calculated as follows (see Appendix A for details of the expression):

$$S_{XX}^+[\omega] = \left(S_{XX}^{s-s}[\omega] + S_{XX}^{c-c}[\omega] + S_{XX}^{s-c}[\omega] + S_{XX}^{c-s}[\omega]\right)/2, \tag{30}$$

$$S_{YY}^-[\omega] = \left(S_{YY}^{s-s}[\omega] + S_{YY}^{c-c}[\omega] - S_{YY}^{s-c}[\omega] - S_{YY}^{c-s}[\omega]\right)/2. \tag{31}$$

From the right-hand side of Eqs. (30) and (31), we note that the quantum correlation noise power spectrum contains four terms. The first two terms represent the self-correlation spectrum of the signal (FWM) field and itself, and the last two terms are derived from the cross-correlation spectra between the signal and FWM fields, which are essential to generating the quantum entanglement.

From Eqs. (30) and (31), it is found that the quantum correlation of the amplitude quadrature sum and the phase quadrature difference are equal, that is, $S_{XX}^+[\omega] = S_{YY}^-[\omega] = S_{sq}[\omega]$. The normalized quantum correlation spectrum in units of dB are defined as $S[\omega] = -10\text{Log}_{10}\left(S_{sq}[\omega]/S_{sn}[\omega]\right)$, in which $S_{sn}[\omega]$ denotes the standard quantum noise limit when both signal and FWM fields are in vacuum states.

The standard quantum noise limit has a value of $S_{sn}[\omega]=1/2$ based on the definition of quadrature components (Eqs. (24–31)).

Figure 5 shows the normalized quantum correlation spectrum of the amplitude (phase) quadrature sum (difference). The parameters of the mechanical resonator we use for simulation are taken from Ref. [42], with the mechanical frequency $\omega_m = 2\pi \times 1.14 \times 10^6$ Hz and the quality factor $Q_m = 1.03 \times 10^9$. The other simulation parameters used are $\kappa = 0.1\,\omega_m$, $\kappa_{ex} = 0.98\,\kappa$, $\omega_0 = 0.95\,\omega_m$, and the environment temperature $T = 1$ K. Note that $\omega = 0$ corresponds to the frequency of the peak intensity gain of the signal (FWM) field, $\Delta_s = -\Delta'_m$ ($-\Delta_s = \Delta'_m$) (Fig. 2(a)).

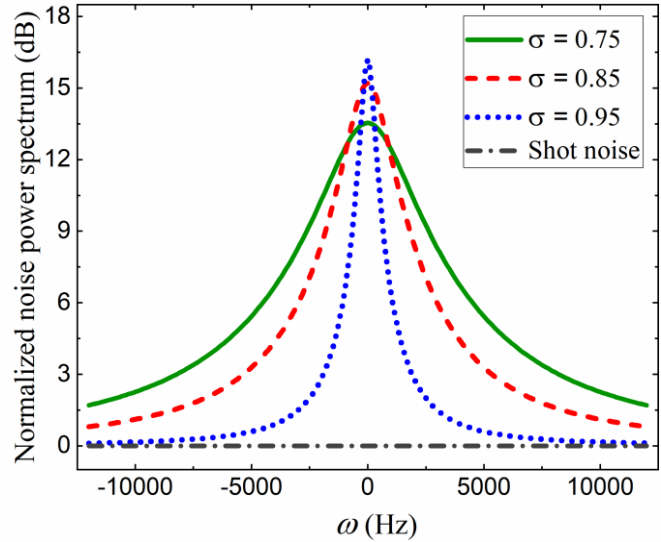

Fig. 5. The normalized noise power spectrum of amplitude (phase) quadrature sum (difference) as a function of analysis frequency under different TMS interaction strength $G_+$. $\sigma = G_+/G_-$ and $G_- = 1.2 \times 10^5$ Hz.

In order to satisfy the weak-coupling condition $G_\pm \ll \kappa$, we fix the coupling strength of the BS interaction $G_- = 1.2 \times 10^5$ Hz and vary the coupling strength of the TMS interaction $G_+$, which is normalized by $G_-$, that is, $\sigma = G_+/G_-$. The quantum entanglement increases with increasing $\sigma$ (or TMS interaction strength $G_+$). For $\sigma = 0.95$, a strong quantum entanglement of 16 dB can be achieved. However, the bandwidth of the noise power spectrum decreases when $\sigma$ ($G_+$) increases, this is due to the effective mechanical damping rate being reduced (see Eq.(13)).

The achievable quantum entanglement is affected by other parameters of the optomechanical system, for instance, the escape efficiency of the optical cavity $\kappa_{ex}/\kappa$. Figure 6 plots the maximum quantum entanglement at the center analysis frequency ($S_{max} = S(0)$) as a function of the escape efficiency of the optical cavity $\kappa_{ex}/\kappa$; the other simulation parameters are the same as those in Fig. 5. It shows that a higher escape efficiency is beneficial to achieving stronger quantum entanglement, and a greater than 9 dB quantum entanglement can be achieved for an escape efficiency $\kappa_{ex}/\kappa$ larger than 0.9. In the range of $\kappa_{ex}/\kappa < 0.9$, the difference of the entanglement for three different ratios of optomechanical coupling strength $\sigma = 0.75, 0.85, 0.95$ are not distinct.

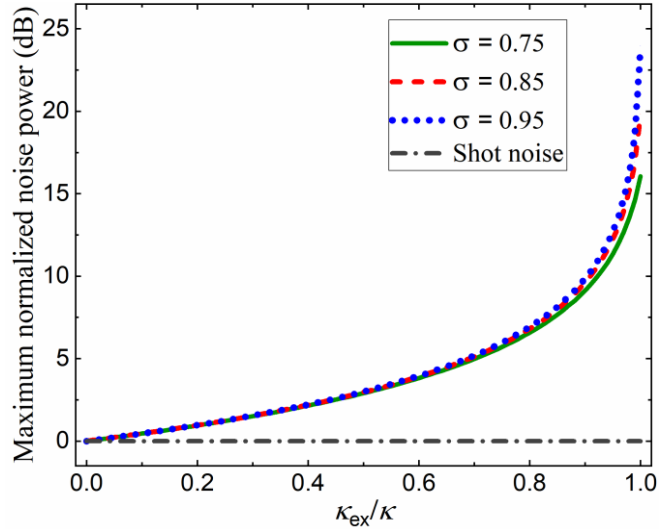

Fig. 6. The maximum normalized noise power $S_{max} = S(0)$ versus the escape efficiency of the optical cavity $\kappa_{ex}/\kappa$. For all cases, $G_- = 1.2 \times 10^5$ Hz.

The achievable quantum entanglement also depends on the environment temperature where the mechanical resonator is located. A large thermal phonon occupation number signifies intense thermal motion of the mechanical resonator, which modulates the optical fields in a non-coherent way and degrades the quantum entanglement of the optical fields. Usually, a mechanical resonator is pre-cooled by a cryogenic cooling system to keep the optomechanical system away from thermal noises. Figure 7 plots the maximum quantum entanglement $S_{max}$ as a function of the environment temperature $T$. The optomechanical coupling strength ratio between the BS interaction $G_-$ and TMS interaction $G_+$ is set to $\sigma = G_+/G_- = 0.95$, and $G_-$ is varied (other parameters are the same as those in Fig. 5).

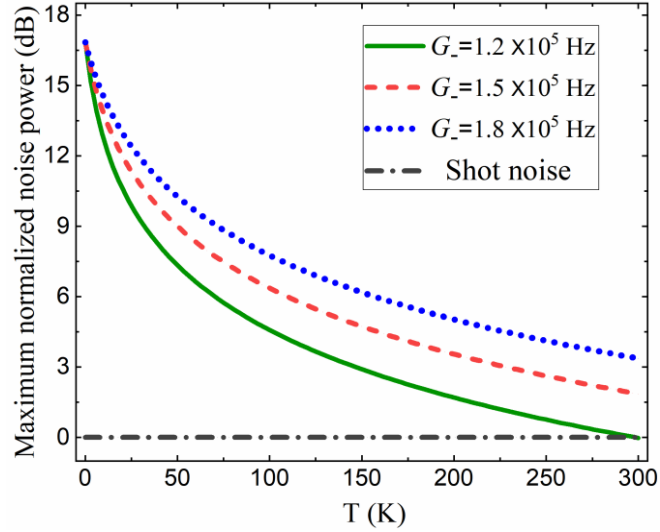

Fig. 7. Maximum normalized noise power $S_{max}$ as a function of the environment temperature $T$ with $\sigma = G_+ / G_- = 0.95$.

For three different BS interaction strengths $G_-$, Fig. 7 shows that the quantum entanglement decreases with the increasing environment temperature, that is, lower thermal phonon occupation is required to achieve a greater quantum entanglement. At finite temperature, the entanglement significantly depends on the BS interaction strength $G_-$ and TMS interaction strength $G_+$. For the coupling strength $G_- = 1.2 \times 10^5$ Hz (green solid line), although the quantum entanglement of 16 dB can be produced at $T = 1$ K, it drops rapidly with increasing temperature and disappears at room temperature ($T = 298$ K). When improving the coupling strength $G_-$ to $1.8 \times 10^5$ Hz (blue dotted line), quantum entanglement beyond 3 dB at room temperature can be obtained. In this case, the thermal phonons' occupation of the mechanical mode $\bar{n}_{th}$ is around $5.4 \times 10^6$. Our scheme is hence robust against thermal mechanical noise, which is particularly important for practical preparation of quantum entanglement.

## V. CONCLUSIONS

In this study, we proposed a scheme to generate strong quantum entanglement by a controllable optomechanical FWM mechanism in a resolved-sideband, and weak-coupling optomechanical system. Firstly, we studied the classical behaviors of the optomechanical FWM process and showed that the frequency response characteristics, including the bandwidth, center frequency, and intensity gain, are tunable by the optomechanical self-energy effect. Then, we investigated the quantum entanglement characteristics between the signal and FWM fields, and found that both the amplitude quadrature sum and phase quadrature difference of the two optical fields are strongly quantum-correlated in proper conditions. We analyzed the key factors that affect the degree of the quantum entanglement and its bandwidth, including the optomechanical coupling strength (BS and TMS interactions) and their ratio, the escape efficiency of the optical cavity, and the initial environment temperature of the mechanical resonator. Quantum entanglements higher than 16 dB and 3 dB can be achieved at cryogenic and room temperatures with the state-of-art optomechanical systems, respectively. The presented scheme provides a promising way for quantum state engineering with low pump power, integrated micro-devices.

## ACKNOWLEDGMENT

This work is supported by the National Key R&D Program of China (Grant No. 2016YFA0301403), the National Natural Science Foundation of China (NSFC) (Grants No. 11774209 and No. 11804208), Key Research and Development Projects of Shanxi Province (201803D121065), and Shanxi 1331KSC.

*yongmin@sxu.edu.cn

## APPENDIX A: THE QUANTUM CORRELATION SPECTRUM

In this appendix, the detailed expressions for the amplitude quadrature sum noise power spectrum $S_{XX}^{+}[\omega]$ and the phase quadrature difference noise power spectrum $S_{YY}^{-}[\omega]$ are derived. From Eq. (17) presented in Section II, the Fourier transform of the conjugate of the fluctuation field $a_{r1}$ is given by

$$a_{r1}^{\dagger}[\omega] = M[\omega]\kappa_{ex}a_{in}[\omega] + M[\omega]\sqrt{\kappa_{ex}\kappa_0}a_v[\omega] + (N[\omega]\kappa_{ex} - 1)a_{in}^{\dagger}[\omega] \\ + N[\omega]\sqrt{\kappa_{ex}\kappa_0}a_v^{\dagger}[\omega] + P[\omega]\sqrt{\kappa_{ex}\gamma_m}\eta[\omega] + Q[\omega]\sqrt{\kappa_{ex}\gamma_m}\eta^{\dagger}[\omega]$$ (A1)

where the related coefficients are

$$M[\omega] = \frac{\chi_c[\omega]\chi_c[\omega]G_-G_+\left(\chi_m[\omega] - \chi_m^*[-\omega]\right)}{\left(1 + \chi_m[\omega]\Sigma[\omega]\right)\left(1 + \chi_m^*[-\omega]\Sigma[\omega]\right)},$$

$$N[\omega] = \frac{\chi_c[\omega]\left\{1 + \chi_c[\omega]\left(G_-^2\chi_m[\omega] - G_+^2\chi_m^*[-\omega]\right)\right\}}{\left(1 + \chi_m[\omega]\Sigma[\omega]\right)\left(1 + \chi_m^*[-\omega]\Sigma[\omega]\right)},$$ (A2)

$$P[\omega] = i\frac{\chi_c[\omega]\chi_m[\omega]G_+}{\left(1 + \chi_m[\omega]\Sigma[\omega]\right)},$$

$$Q[\omega] = i\frac{\chi_c[\omega]\chi_m^*[-\omega]G_-}{\left(1 + \chi_m^*[-\omega]\Sigma[\omega]\right)}.$$

To derive the quadrature noise power spectrum, the following correlation relations are exploited, including the vacuum noises correlation of optical fields, and thermal noise correlation of the mechanical mode in the frequency domain:

$$\begin{aligned}
\langle a_{in,v}[\omega] a_{in,v}^\dagger[\omega']\rangle &= \delta(\omega+\omega'), \\
\langle a_{in,v}^\dagger[\omega] a_{in,v}[\omega']\rangle &= 0, \\
\langle \eta[\omega]\eta^\dagger[\omega']\rangle &= (\bar{n}_{th}+1)\delta(\omega+\omega'), \\
\langle \eta^\dagger[\omega]\eta[\omega']\rangle &= \bar{n}_{th}\delta(\omega+\omega'),
\end{aligned} \quad (A3)$$

where $\bar{n}_{th}$ denotes the initial thermal phonons' occupation of the mechanical mode; it is proportional to the environment temperature where the mechanical resonator is located. Using Eqs. (11), (17), and (A1–A3), we obtain the final results of the four terms on the right-hand side of Eqs. (30) and (31):

$$S_{XX}^{s-s}[\omega] = S_{YY}^{s-s}[\omega] = 0.5 \left\{ \begin{array}{l} (A[\omega_1]\kappa_{ex}-1)(N[-\omega_1]\kappa_{ex}-1) + M[\omega_2]B[-\omega_2]\kappa_{ex}^2 \\ +(A[\omega_1]N[-\omega_1]+M[\omega_2]B[-\omega_2])\kappa_{ex}\kappa_0 \\ +\kappa_{ex}\gamma_m \left[ \begin{array}{l} (C[\omega_1]Q[-\omega_1]+P[\omega_2]Q[-\omega_2])(\bar{n}_{th}+1) \\ +(D[\omega_1]P[-\omega_1]+Q[\omega_2]C[-\omega_2])\bar{n}_{th} \end{array} \right] \end{array} \right\}, \quad (A4)$$

$$S_{XX}^{c-c}[\omega] = S_{YY}^{c-c}[\omega] = 0.5 \left\{ \begin{array}{l} (A[\omega_2]\kappa_{ex}-1)(N[-\omega_2]\kappa_{ex}-1) + M[\omega_1]B[-\omega_1]\kappa_{ex}^2 \\ +(A[\omega_2]N[-\omega_2]+M[\omega_1]B[-\omega_1])\kappa_{ex}\kappa_0 \\ +\kappa_{ex}\gamma_m \left[ \begin{array}{l} (C[\omega_2]P[-\omega_2]+P[\omega_1]Q[-\omega_1])(\bar{n}_{th}+1) \\ +(D[\omega_2]P[-\omega_2]+Q[\omega_1]C[-\omega_1])\bar{n}_{th} \end{array} \right] \end{array} \right\}, \quad (A5)$$

$$S_{XX}^{s-c}[\omega] = -S_{YY}^{s-c}[\omega] = 0.5 \left\{ \begin{array}{l} \left[(A[\omega_1]\kappa_{ex}-1)B[-\omega_1]+M[\omega_2](N[-\omega_2]\kappa_{ex}-1)\right]\kappa_{ex} \\ +(A[\omega_1]B[-\omega_1]+M[\omega_2]N[-\omega_2])\kappa_{ex}\kappa_0 \\ +\kappa_{ex}\gamma_m \left[ \begin{array}{l} (C[\omega_1]D[-\omega_1]+P[\omega_2]Q[-\omega_2])(\bar{n}_{th}+1) \\ +(D[\omega_1]C[-\omega_1]+Q[\omega_2]P[-\omega_2])\bar{n}_{th} \end{array} \right] \end{array} \right\}, \quad (A6)$$

$$S_{XX}^{c-s}[\omega] = -S_{YY}^{c-s}[\omega] = 0.5 \left\{ \begin{array}{l} \left[(A[\omega_2]\kappa_{ex}-1)B[-\omega_2]+M[\omega_1](N[-\omega_1]\kappa_{ex}-1)\right]\kappa_{ex} \\ +(A[\omega_2]B[-\omega_2]+M[\omega_1]N[-\omega_1])\kappa_{ex}\kappa_0 \\ +\kappa_{ex}\gamma_m \left[ \begin{array}{l} (C[\omega_2]D[-\omega_2]+P[\omega_1]Q[-\omega_1])(\bar{n}_{th}+1) \\ +(D[\omega_2]C[-\omega_2]+Q[\omega_1]P[-\omega_1])\bar{n}_{th} \end{array} \right] \end{array} \right\}, \quad (A7)$$

where $\omega_1 = \omega - \Delta_s$ and $\omega_2 = \omega + \Delta_s$.